\documentclass{article}     		
\title{\textbf{A magnetic guide for cold atoms}}
\author{\normalsize{James A. Richmond, S\'\i le  Nic 
Chormaic, Benjamin P. Cantwell and Geoffrey I. Opat}\\
\normalsize{\textit{School of Physics, University of Melbourne, Parkville,
Victoria 3052, Australia}}}

\usepackage{amsmath}
\usepackage{latexsym}

\begin{document}            
\maketitle 

\normalsize
\begin{abstract}
We propose a novel method for guiding cold, 
neutral atoms using static magnetic fields.  A theoretical study of the  
magnetic field produced by a  tube consisting of two identical,
 interwound 
solenoids carrying equal but opposite currents is presented. 
This field is almost zero throughout the centre of the tube, but it 
increases with exponential rapidity as one approaches 
the walls formed by the current carrying wires.  
Hence, cold atoms passing through the tube may be reflected by 
 magnetic mirror  effects near the walls. 
Applying this  technique to a free-falling cloud of magneto-optically
cooled caesium atoms we hope to construct atomic guides  to
facilitate the manipulation of cold atomic beams. 
\end{abstract}
\vspace{5mm}
\begin{center}\textbf{1. Introduction} \end{center}
 
During recent years, with the advent of easily produced cold atomic 
sources, 
significant interest has been displayed in developing new 
and useful methods for manipulating and controlling  cooled atoms. 
This has resulted in the realisation of many atom optical elements such
as beam-splitters, mirrors and diffraction gratings.  For a recent 
review on some of the progress in this field see, for example,
ref. \cite{BERM}.  

The majority of atom optical components rely on 
the effects  either of magnetic fields or of light fields
on neutral atoms. One clear example of this is shown by 
the different types of atomic mirrors which have successfully 
demonstrated atomic reflections.  Permanent 
magnets forming an array of alternating magnetic poles 
\cite{ OPAT1, SID1}, 
ferromagnetic surface mirrors made from magnetic audio-tape
\cite{HIND} and  current carrying wires arranged so that 
adjacent wires have equal but opposite currents \cite{WAY, LAU}
 all rely on  the interaction of 
the permanent atomic magnetic dipole moment and a non-uniform magnetic field.  
An atom in an appropriate magnetic state is repelled by the  
magnetic field gradient.  
Specular reflection of  thermal \cite{BALY}
and laser-cooled \cite{CHU} atoms
from the repulsive potential of a blue-detuned evanescent light wave has also
been observed.  This method is based on the  interaction between 
 an induced atomic electric dipole moment and the non-uniform
field of an evanescent light wave \cite{COOK}. 

A natural progression from the successful realisation of atomic mirrors 
has been the development of atomic guides based on the different
reflection mechanisms.  Until recently, guides 
relying on atomic reflections by evanescent 
light fields within a hollow optical fibre [10 - 12]
have been the main candidates in this field.  Several other guiding
methods  have been proposed [13 - 16].  

 In this paper we propose an
 atomic guide based on the reflection of cold atoms 
by current carrying wires.  A suitable magnetic field gradient can be
produced by two interwound solenoids carrying equal but opposite
currents. 
We expect that such a guide will be more 
adaptable to different experimental studies than
 the transfer technique used in \cite{JILA}, mainly due to the fact
that it can be positioned very close to the MOT by placing it
\text{inside} the vacuum chamber. Losses due to 
trap expansion are therefore minimised.  When the MOT is operating
the current through the guide can be switched off and there will
be no distorting effects on the trap.  This is in stark contrast to guide
designs based on permanent magnets. The following sections present the concept 
behind the magnetic 
guide and the magnetic field configuration to be used.

\vspace{5mm}
\begin{center}\textbf{2. The magnetic guide} \end{center} 
Let us consider the interaction between atoms
with an atomic magnetic dipole moment,
\mbox{\boldmath${\mu}$}, and a static inhomogeneous magnetic field,
\mbox{\boldmath${B}$}.  In this case, the interaction potential 
is position dependent. We shall assume adiabaticity for the particular
case of slowly moving, laser-cooled atoms.  
In other words, we assume that the projection of the atomic angular
momentum onto the magnetic field direction remains constant during
the atom-magnetic field interaction.   The 
 force on the  laser-cooled atoms in the presence of the
\mbox{\boldmath${B}$}-field is given by
\begin{equation}
\mbox{\boldmath $F$} = \nabla(\mbox{\boldmath $\mu \cdot B$}) 
	= -\mathit{m g_{F} \mu_{B} \nabla B},
\label{eq:force}
\end{equation}
where $m$ is the magnetic quantum number, $g_{F}$ is the Land\'{e} 
$g$-factor and $\mu_{B}$ is the Bohr magneton.  From eq.~(\ref{eq:force})
we see that  the atoms
will be repelled by an increasing magnetic field gradient
when the product $m g_{F}$ is positive.  This
is  the main principle behind magnetic mirrors. 

As has been mentioned in the previous section, an effective and efficient 
type of mirror for cold, neutral atoms can be produced by reflection 
from a magnetic surface.  Free-falling, laser-cooled caesium atoms
pumped into the  $m = +4$ hyperfine state have already been reflected 
from an atom mirror
consisting of an array of current carrying wires, with the current 
in adjacent wires being equal but opposite in direction \cite{LAU}
as shown in fig.~1(a).  

 If we now consider an extension of this study, we can 
effectively imagine bending the wire array mentioned
previously, in order to form a tube through
which the atoms can pass.  Each wire from the array now forms a 
current loop and this is represented schematically in fig.~1(b).
The current in each loop is equal and the currents in two adjacent loops
are in opposite directions. This design  results in the 
atoms being reflected irrespective of the path they follow
through the tube.     The reflections cause
the atoms to be trapped inside the tube rather than
being absorbed on the walls. Therefore, the tube 
forms an atomic guide.  A tube formed by adjacent
current loops, as shown in fig.~1(b), 
can be approximated by two interwound solenoids.  
In the next section we will discuss 
the magnetic field of such an arrangement and show that it provides us with 
appropriate conditions for constructing an atomic guide. 
Theoretical magnetic field plots for our particular
experimental conditions will also be presented.

\vspace{5mm}
\begin{center}\textbf{3. Magnetic fields of helical solenoids} \end{center}
An expression for the magnetic field of an infinitely long solenoid
can be found by exploiting its helical symmetries. Although these 
calculations
are tedious to perform, the final result shows a definite magnetic 
field structure
beyond that obtained via the usual elementary calculation based
on Amp\`{e}re's circuital law and the current sheet approximation. 
 
Let us consider a solenoid of radius $R$ and let $a$ be the distance
between adjacent wires in the $z$-direction, \textit{i.e.} along
the long axis of the solenoid.  Note that $a$ is the
reciprocal of the number of turns of wire per unit length.  The path
of the wire wound in the solenoidal coil is given by $r=R$, 
$\phi = 2 \pi \nu$ and $z = z_0 + a \nu$ in cylindrical coordinates.  
$\nu$ is a continuous parameter describing the number of turns of wire
as counted from an arbitrary starting point, $z=z_0$.  
We are 
particularly interested in the magnetic field inside the solenoid
(\textit{i.e.} where $r < R$).  It can be shown that the 
magnetic field components, $(B_r, B_\phi, B_z)$, 
are given by the following expressions \cite{JAMES}: 
\begin{eqnarray}
B_r &=& \frac{\mu_0 I 4 \pi R}{a^2}
     \sum_{n=1}^{\infty} n \sin \left[
      n\left(\phi - \frac{2 \pi}{a}\left(z-z_0\right)\right)\right] 
	{K_n}'(\rho_0){I_n}'(\rho), \label{eq:bfield1}\\
B_\phi &=& \frac{\mu_0 I 2 R}{ar}
      \sum_{n=1}^{\infty} n \cos \left[
      n\left(\phi - \frac{2 \pi}{a}\left(z-z_0\right)\right)\right]
	{K_n}'(\rho_0)I_n(\rho), \label{eq:bfield2}\\
B_z &=& \frac{\mu_0 I}{a} -\frac{\mu_0 I 4 \pi R}{a^2}
      \sum_{n=1}^{\infty} n \cos \left[
	n\left(\phi - \frac{2 \pi}{a}\left(z-z_0\right)\right)\right]
	{K_n}'(\rho_0)I_n(\rho).
\label{eq:bfield3}
\end{eqnarray}
By definition, $\rho_0~\equiv~2\pi nR/a$ and $\rho~\equiv~2\pi nr/a$.
$I$ is the current in the solenoid and  $z_0$ 
is an arbitrary point along the $z$-axis through which the wire passes.
The $K_n(\rho)$ and the $I_n(\rho)$ are modified Bessel functions. 
The primes on the $K_n$
and $I_n$ indicate differentiation with respect to $\rho$. 

In order to construct an atomic mirror, we know that we need an 
array of wires such that adjacent wires carry equal but opposite
current.  Extending this to the idea of an atomic guide it
seems reasonable  that two interwound solenoids will give us the correct
magnetic field gradient to produce atomic reflections in 
all directions (\textit{c.f.} fig. 1).  Each solenoid
is a single wire solenoid of radius $R$ and  distance $a$ 
between adjacent turns of wire.  The two solenoids must also carry 
equal but opposite currents.  This is treated mathematically by 
giving each solenoid a different reference point $(0,0,z_0)$ 
through which the wires pass.  For one solenoid we take the current
to be $I$ and $z_0~=~0$ and for the second coil the current is $-I$ and
$z_0~=~a/2$.  Using eqs.~(\ref{eq:bfield1}), (\ref{eq:bfield2}) and
(\ref{eq:bfield3}) we can easily obtain the 
solutions for the magnetic field  \mbox{\boldmath${B}$} produced
by the double wound solenoid.  We are again interested in the field
\emph{inside} the tube formed by the two solenoids (\textit{i.e.} 
where $r < R$).  The magnetic field components are given by
\begin{eqnarray}
B_r &=& \frac{\mu_0 I 8 \pi R}{a^2}
      \sum_{n=0}^{\infty} \left(2n+1\right) \sin \left[
      \left(2n+1 \right)\left(\phi - \frac{2\pi z}{a}\right)\right]
      \nonumber\\ &~&~~\times
      {K_{2n+1}}'\left(\left(2n+1\right) \frac{2 \pi R}{a}\right)
      {I_{2n+1}}'\left(\left(2n+1\right) \frac{2\pi r}{a}\right), 
 	\label{eq:double1}\\
B_\phi &=& \frac{\mu_0 I 4 R}{ar}
      \sum_{n=0}^{\infty} \left(2n+1\right) \cos \left[
      \left(2n+1\right)\left(\phi - \frac{2\pi z}{a}\right)\right]
      \nonumber\\ &~&~~\times
      {K_{2n+1}}'\left(\left(2n+1\right) \frac{2 \pi R}{a}\right)
      I_{2n+1}\left(\left(2n+1\right) \frac{2\pi r}{a} \right), 
	\label{eq:double2}
\end{eqnarray}
\begin{eqnarray}
B_z &=& -\frac{\mu_0 I 8 \pi R}{a^2}
      \sum_{n=0}^{\infty} (2n+1) \cos \left[
      \left(2n+1\right)\left(\phi - \frac{2\pi z}{a}\right)\right]
      \nonumber\\ &~&~~\times
      {K_{2n+1}}'\left(\left(2n+1\right) \frac{2 \pi R}{a}\right)
      I_{2n+1}\left(\left(2n+1\right) \frac{2\pi r}{a}\right),
\label{eq:double3}
\end{eqnarray}
where the various symbols represent the same quantities 
as in eqs.~(\ref{eq:bfield1}), (\ref{eq:bfield2}) and (\ref{eq:bfield3}).

\vspace{5mm}
\begin{center}\textbf{4. Magnetic field profiles} \end{center}
Figure 2(a)  shows a  vertical slice (\textit{i.e.} 
along the \textit{z}-axis)  of the magnetic tube formed
by the two solenoids.  The magnetic field is calculated using 
eqs.~(\ref{eq:double1}), (\ref{eq:double2}) and (\ref{eq:double3}) and
considering the first 10 terms of the Fourier series only.   
In this example,
we have assumed that $a/R=1.0$ and that a current of 0.5Amps flows
through each solenoid. 
The field inside the tube is 
close to zero throughout the central region.  
This is identical to the result
predicted using Amp\`ere's circuital law for two interwound solenoids. 
However, there is a large  magnetic field gradient
as one approaches the wires.  Figure 2(b) is a 
horizontal view of the 
magnetic field ($z=0$ plane), using the same parameters as in
 fig.~2(a).  
Free-falling, cold atoms enter the tube normal to the plane of the diagram.  
Any off-axis
atoms will  experience a steep field gradient near the walls
and will
be reflected back towards the centre.  
It is important that the magnetic force  is large enough
so that the atoms in the ballistically expanding cloud never quite reach
the walls of the tube.  This ensures
that they will be reflected \emph{before} striking the wires and thus,
a high transport efficiency can be obtained for the guide, due to 
the low loss in flux from absorption.  

Similar plots of the 
\mbox{\boldmath${B}$}-field are shown in fig. 3.  In this case, the 
solenoids are more tightly wound than in fig. 2 so that $a/R = 0.1$. 
Again, we assume a current of 0.5Amps. 
Using these parameters we see that the  magnetic field remains 
essentially uniform over a larger region than that shown in fig.
2.  However, the magnetic field gradient
 is steeper and a higher efficiency of reflection should
be obtained.  

\vspace{5mm}
\begin{center}\textbf{5. Design criteria} \end{center}
Ultra-high vacuum compatible, Kapton-coated wires of maximum diameter 0.86mm
are  suitable for 
making the magnetic guide.  Considering an atom guide of 
radius 2.5mm this yields $a/R$ = 0.69, which lies within the range we 
have studied.    
A typical transverse velocity for trapped Cs atoms
is $\pm$6cm/s.   Let us assume that the entrance to the atom guide is located 
1.5cm beneath the centre of the trapped cloud of atoms and the guide
 is 3.8cm long.  Once the trap is switched off, we predict that 67\% of 
the free-falling atoms
will enter the guide.  45\% of these  atoms will strike the walls of the guide
and be reflected.  The other 55\% of atoms will fall straight through. 
A longer guide would reflect a higher percentage of atoms.  
Figure 4 is a 
plot of the current needed to reflect atoms from the walls of the guide
 versus the distance from the centre of the tube at which the atoms will
be reflected.   From 
this graph we see that atoms  close to the walls (2.2mm from the centre) 
require a current of approximately 0.1Amps to be reflected. Our practical
limit of 0.5Amps is therefore largely sufficient.   

\vspace{5mm}
\begin{center}\textbf{6. Conclusion} \end{center}
We have presented a new scheme for constructing atomic guides, 
based on the reflection of cold atoms from current carrying wires.  Magnetic
mirrors consisting of a current carrying wire array have already
been realised \cite{LAU}.  Our guide scheme is an extension of this 
technique. A 
suitable magnetic field configuration for the guide consists
of two interwound solenoids.  This magnetic atom guide has
 a zero-field region along its axis and 
a steep field gradient as one approaches the walls radially 
from the centre.   In order to avoid 
non-adiabatic spin-flips in the zero 
region of the \mbox{\boldmath${B}$}-field, it may be necessary to add
a small bias field to the overall configuration by unbalancing the 
currents in the two solenoids by a small amount.

This magnetic guide design has a distinct advantage over
other guiding schemes \cite{Sav, REN, PFAU} since it doesn't 
require additional lasers and
 is, therefore,  very cost effective and simple in construction.  Because
it is based on current carrying wires it can be switched on and off 
as required.  Therefore 
its entrance can be  positioned close to the MOT.  
Any constraints on the distance between
the MOT and the atom guide arise from the  diameter of 
the trapping laser beams and the radius of the guide itself.
 Therefore, by a judicious choice of these two parameters, 
the free-falling cloud of cold atoms should not expand appreciably before
entering the magnetic tube. The entire experiment should be switched so 
that the MOT is on when the
magnetic guide is off and vice versa and the guide should be 
switched on after the atoms have entered it.  This avoids any
 problems that may arise
from heat dissipation and distorting effects of the magnetic fields.

\vspace{5mm}
\noindent\textbf{Acknowledgements} The authors wish to thank Andrei 
Sidorov, Tony Klein and Peter Hannaford for useful discussions.  
JAR would
like to acknowledge the assistance of an APA and 
the Pam Todd Scholarship from St. Hilda's College. SNC is
supported by a GIRD (AMIRA) grant. BPC is funded by an MRS. 

\pagebreak
\normalsize
\noindent Fig. 1: (a) Magnetic mirror formed using a 
continuous array of current carrying wires.  The black arrows 
represent the direction of the current in each wire; 
(b) The same as (a) only each wire is folded 
through $360^o$, thus forming a tube of 
current loops. 

\vspace{5mm}
\noindent Fig. 2: Magnetic field of a double wound solenoid for 
$a/R = 1.0$ and a current $I=0.5$A. 
(a) Field magnitude for a vertical slice through
the magnetic tube ($y = 0$ plane); (b) Field magnitude
for a horizontal slice through the magnetic tube ($z=0$ plane). 
 
\vspace{5mm}
\noindent Fig. 3: Magnetic field of a double wound solenoid for 
$a/R = 0.1$ and a current $I=0.5$A.  
(a) Field magnitude for a vertical slice through
the magnetic tube ($y = 0$ plane); (b) Field magnitude
for a horizontal slice through the magnetic tube ($z=0$ plane).

\vspace{5mm}
\noindent Fig. 4: Plot of current through the solenoids versus distance from
the centre at which an atom will be reflected.  $a/R = 0.69$. 
It is assumed that the atoms
are in the $m = +4$ state and have a transverse velocity of $\pm$6cm/s.  
Atoms close to 
the walls of the guide will be relected for currents $\leq 0.5$Amps.

\end{document}